\documentclass{mem}
\usepackage{natbib}
\usepackage{txfonts}
\usepackage{graphicx}
\usepackage{flushend}
\usepackage{epsfig}
\usepackage{epstopdf}
\usepackage[a4paper,breaklinks,dvipdfm]{hyperref}
\idline{75}{282}
\begin{document}
\def\teff{$T\rm_{eff }$}
\def\kms{$\mathrm {km s}^{-1}$}

\def\eze{$E_{\rm 0}$ }
\def\eppo{$E_{\rm p}$ }
\def\epi{\ensuremath{E_{\rm p,i}}}
\def\eiso{\ensuremath{E_{\rm iso}}}
\def\ega{\ensuremath{E_{\rm jet}}}
\def\eiso{\ensuremath{E_{\rm iso}}}
\def\liso{\ensuremath{L_{\rm iso}}}
\def\lpiso{\ensuremath{L_{\rm p,iso}}}
\newcommand{\epeiso}{$E_{\rm peak}-E_{\rm iso}$}
\newcommand{\epliso}{$E_{\rm peak}-L_{\rm iso}$}
\newcommand{\ep}{$E_{\rm peak}$}
%\newcommand{\epo}{$E^{\rm obs}_{\rm peak}$}
%r
%\newcommand{\eiso}{$E_{\rm iso}$}
%\newcommand{\liso}{$L_{\rm iso}$}
\providecommand{\url}[1]{\texttt{#1}}
\providecommand{\href}[2]{#2}
\providecommand{\path}[1]{#1}
\providecommand{\DOIprefix}{doi:}
\providecommand{\ArXivprefix}{arXiv:}
\providecommand{\URLprefix}{URL: }
\providecommand{\Pubmedprefix}{pmid:}
\providecommand{\doi}[1]{\href{http://dx.doi.org/#1}{\path{#1}}}
\providecommand{\Pubmed}[1]{\href{pmid:#1}{\path{#1}}}
\providecommand{\bibinfo}[2]{#2}

\def\ega{\ensuremath{E_{\gamma}}}
\def\epeiso{$E_{\rm p,i}$ -- $E_{\rm iso}$}
\def\nufnu{$\nu F_{\nu}$ }
\def\epega{$E_{\rm p,i}$ -- $E_{\gamma}$ }
\def\epeisotb{$E_{\rm p,i}$ -- $E_{\rm iso}$ -- $t_{\rm b}$ }
\def\sext{$\sigma_{\rm ext}$}
\def\sax{ Beppo SAX }
\def\swift{Swift}
\def\fermi{Fermi}
\def\konus{Konus}
\def\suz{Suzaku}

\title{High redshift constraints on dark energy models from the $E_{\rm p,i}$ -- $E_{\rm iso}$  correlation in GRBs.}

   \subtitle{}

\author{
Marek Demianski\inst{1,2} 
Ester Piedipalumbo\inst{3,4}
Disha Sawant\inst{5,6}   
Lorenzo Amati\inst{7}}

\institute{
Institute for Theoretical Physics, University of Warsaw, Pasteura 5, 02-093 Warsaw, Poland.
\and
Department of Astronomy, Williams College, Williamstown, MA 01267, USA.
\and
Dipartimento di Fisica, Universit\`{a} degli Studi di Napoli Federico II, Compl. Univ. Monte S. Angelo, 80126 Naples, Italy
\email{ester@na.infn.it}. 
\and
I.N.F.N., Sez. di Napoli, Compl. Univ. Monte S. Angelo, Edificio 6, via Cinthia, 80126 - Napoli, Italy.
\and
Dipartimento di Fisica e Scienze della Terra, Universit\`{a} degli Studi di Ferrara
\and
Department of Physics, University of Nice Sophia Antipolis, Parc Valrose 06034, Nice Cedex2, France
\and
INAF-IASF, Sezione di Bologna, via Gobetti 101, 40129 Bologna, Italy
}

\authorrunning{Demianski et al }

\titlerunning{Cosmology with the $E_{\rm p,i}$ -- $E_{\rm iso}$  correlation }

\abstract{
 Here we test  different models of dark energy beyond the standard cosmological
constant scenario.  We start considering the CPL parameterization of the equation of state (EOS), then we consider a dark energy scalar field (Quintessense). Finally we consider models  with dark energy at early times (EDE).  Our analysis is based on the Union2 type Ia supernovae data set,  a Gamma Ray Bursts
(GRBs) Hubble diagram, a set of 28 independent measurements of the Hubble parameter, some baryon acoustic oscillations
(BAO) measurements. We performed a statistical analysis and explore the probability distributions of the cosmological parameters for each of the competing models. To build up their own regions of confidence,  we maximize some appropriate likelihood functions using the Markov chain Monte Carlo (MCMC)
method.  Our analysis indicates that the EDE and the scalar field quintessence are slightly favored by the present data. Moreover, the GRBs Hubble diagram alone is able to
set the transition region from the decelerated to the accelerated expansion of the Universe in all the tested models.
Perspectives for improvements in the field with the THESEUS mission are also described. 

}
\maketitle{}

\section{Introduction}

Starting at the end of the 1990s, observations of high-redshift supernovae of type Ia (SNIa) revealed the current
accelerated expansion of the Universe  \citep[see e.g.][]{perl98, per+al99, Riess, Riess07,SNLS,Union2}, 
which is driven by the so called dark energy. The so far proposed models of dark energy  range from a non-zero 
cosmological constant \citep[see for instance][]{carroll01}, to a potential energy of some not
yet discovered scalar field \citep[see for instance][]{SF}, or effects connected with the inhomogeneous distribution
of matter and averaging procedures \citep[see for instance][]{clark}. In these last two cases the equation of state, 
EOS, depends on the redshift $z$. To probe the dynamical evolution of dark energy we consider different
competitive cosmological scenarios:
\begin{itemize}
\item[i)] an EOS empirically parametrized,
\item[ii)] a scalar field model for  dark energy,
\item[iii)] an early time dark energy model.
\end{itemize}
In our high-redshift  investigation, extended beyond the supernova type Ia (SNIa) Hubble diagram, we use the Union2
SNIa data set, the gamma-ray burst (GRB) Hubble diagram, constructed by calibrating the correlation between the peak
photon energy, $E_{\mathrm{p, i}}$, and the isotropic equivalent radiated energy, $ E_{\mathrm{iso}}$ \cite{MGRB2}.
Here we take into account  possible redshift evolution effects in the coefficients of this correlation, assuming that
they can be modeled  through power low terms. We consider also a sample of 28 measurements of the Hubble parameter,
compiled in \cite{farooqb},  Gaussian priors on the distance from the baryon acoustic oscillations (BAO), and the
Hubble constant $h$. Our statistical analysis is based on Monte Carlo Markov Chain (MCMC) simulations to simultaneously compute the full
probability density functions (PDFs) of all the parameters of interest. 
\section{Competitive Dark Energy models}
Here we are  looking for some dynamical field that is generating an effective negative pressure. Moreover this could instead be indicating that the Copernican principle cannot be applied at certain scales, and that radial inhomogeneity could mimic the accelerated expansion. Within the
Friedman-Lemaitre-Robertson- Walker (FLRW) paradigm, all possibilities can be characterized by the dark energy EOS, $w(z)$. A prior task of observational cosmology  is to search for
evidence for $w(z) \neq -1$.  If we assume that the dark energy evolves, the importance
of its equation of state is significant and it determines the Hubble function $H(z)$, and any derivation of it is
needed to obtain the observable quantities. Actually it turns out that:
\begin{eqnarray}\label{heos}
  H(z,{\mathrm \theta}) &=& H_0 \sqrt{(1-\Omega_m) g(z, {\mathrm \theta})+\Omega_m (z+1)^3} \nonumber 
\end{eqnarray}
 where $g(z, {\mathrm \theta})=\frac{\rho_{de}(z)}{\rho_{de}(0)}=\exp^{3 \int_0^z \frac{w(x,{\mathrm \theta})+1}{x+1} \, dx}$,  $w(z,{\mathrm \theta})$ is any dynamical form of the dark energy EOS, and ${\mathrm \theta}=(\theta_1, \theta_1..,\theta_n)$ are the EOS  parameters.
In  the  Chevalier-Polarski Linder (CPL) model \cite{cpl1,cpl2},
the dark energy EOS given by
\begin{equation}
w(z) =w_0 + w_{1} z (1 + z)^{-1} \,,
\label{cpleos}
\end{equation}

\subsection{A scalar field quintessence model}\label{scalar}
 In this section the possible physical realization
of dark energy  is a cosmic scalar field, $\varphi$, minimally coupled to the usual matter action. Here we take into account the specific class of
exponential--type potential; in particular we consider an exponential potential for which  general exact solutions of
the Friedman equations are known \cite{MEC11,EPGC}. Assuming that  $\varphi$ is minimally coupled to gravity, the
cosmological equations are written as
\begin{equation}
H^{2} = \frac{8\pi G}{3}(\rho _{M} + \rho _{\varphi})\,, \label{eq1}
\end{equation}
\begin{equation}
\dot{H} + H^{2} = -\frac{4\pi G}{3}(\rho _{M} + \rho _{\varphi} + 3(p_{M} + p_{\varphi}))\,, \label{eq2}
\end{equation}
\begin{equation}
\ddot{\varphi} + 3H\dot{\varphi} + V^{\prime}(\varphi) = 0\,. \label{eq3}
\end{equation}
Here
\begin{equation}
\rho_{\varphi} \equiv \frac{1}{2}{\dot{\varphi}}^2 + V(\varphi)\,,\,\,\,\,\,\,\, p_{\varphi} \equiv
\frac{1}{2}{\dot{\varphi}}^2 - V(\varphi)\,, \label{eq5}
\end{equation}
and
\begin{equation}
w_{\varphi} \equiv \frac{{\dot{\varphi}}^2 - 2V(\varphi)}{{\dot{\varphi}}^2 + 2V(\varphi)}\,. \label{eq6}
\end{equation}

We consider the potential
\begin{equation}\label{scal1}
V(\varphi) \propto \exp\left\{ -\sqrt{3\over 2}\varphi\right\}\,,
\end{equation}
for which the general exact solution exists \citep{MEC11,EPGC}.

\subsection{Early Dark Energy}\label{ede}
In this section we consider a model characterized by a non negligible amount of dark energy at early
times:  these models  are connected to the existence of scaling  or attractor-like solutions, and they naturally predict a non-vanishing dark
energy fraction  of the total energy at early stages, $\Omega_{e}$, which should be substantially smaller than its
present value.  Following \cite{doran06, pettorino13} we use a
parametrized representation of the  dark energy density fraction, $\Omega_{DE}$, which depends on the present matter
fraction, $\Omega_m$, the early dark energy density fraction, $\Omega_e$ , and the present dark energy equation of
state $w_0$:
\begin{eqnarray*}\label{edep}
&&\Omega_{DE}(z,\Omega_m, \Omega_e, w_0)=\nonumber\\&&\frac{\Omega _e \left(-\left(1-(z+1)^{3 w_0}\right)\right)-\Omega _m+1}{\Omega _m(z+1)^{-3 w_0}-\Omega _m+1}\nonumber\\
   && +\Omega _e \left(1-(z+1)^{3 w_0}\right)\,.
\end{eqnarray*}
It turns out that the Hubble function takes the form:
 \begin{eqnarray}\label{Hede}
&&H^{2}(z,\Omega_m, \Omega_e, w_0,\Omega _{\gamma }, N_{eff })=\nonumber\\&&\Omega_{DE}(z,\Omega_m, \Omega_e, w_0)+\nonumber \\
&&+ (z+1)^3\Omega _m+\nonumber\\&&(z+1)^4 \Omega _{\gamma }\left( \frac{7}{8}\left(\frac{4}{11}\right)^{\frac{4}{3}} N_{eff }+1\right).
 \end{eqnarray}
 Here $N_{eff}=3$ for three standard model neutrinos that were thermalized in the early Universe and decoupled well before electron-positron annihilation.

\section{Observational data}
In our approach we use measurements on SNIa and
GRB Hubble diagram,  distance data from the BAO, and a list of
$28$ $H(z)$ measurements, compiled in \cite{farooqb}.

\subsection{Supernovae Ia}
 SNIa observations gave the first strong indication of the recent accelerating
expansion of the Universe. First results of the SNIa teams were
published by  \cite{Riess} and \cite{per+al99}. Here we consider
the recently updated Supernovae Cosmology Project Union 2.1
compilation \cite{Union2.1}, which is an update of the original
Union compilation and contains $580$ SNIa, spanning the redshift
range ($0.015 \le z \le 1.4$). We compare the theoretically
predicted distance modulus $\mu(z)$ with the observed one
through a Bayesian approach, based on the definition of the
distance modulus in different cosmological models:

\begin{equation}
\mu(z_{j}) = 5 \log_{10} ( D_{L}(z_{j}, \{\theta_{i}\}) )+\mu_0\,,
\end{equation}
where $D_{L}(z_{j}, \{\theta_{i}\})$ is the Hubble free
luminosity distance, and $\theta_{i}$ indicates the set of
parameters that appear in different dark energy equations of
state considered in our analysis. The parameter $\mu_{0}$
encodes the Hubble constant and the absolute magnitude $M$.
Given the heterogeneous origin of the Union data set, we used an
alternative version of the $\chi^2$:
\begin{equation}
\label{eq: sn_chi_mod}
\tilde{\chi}^{2}_{\mathrm{SN}}(\{\theta_{i}\}) = c_{1} -
\frac{c^{2}_{2}}{c_{3}}\,,
\end{equation}
where
\begin{equation}
c_{1} = \sum^{{\cal{N}}_{SNIa}}_{j = 1} \frac{(\mu(z_{j}; \mu_{0}=0,
\{\theta_{i})\} -
\mu_{obs}(z_{j}))^{2}}{\sigma^{2}_{\mathrm{\mu},j}}\, ,
\end{equation}
\begin{equation}
c_{2} = \sum^{{\cal{N}}_{SNIa}}_{j = 1} \frac{(\mu(z_{j}; \mu_{0}=0,
\{\theta_{i})\} -
\mu_{obs}(z_{j}))}{\sigma^{2}_{\mathrm{\mu},j}}\, ,
\end{equation}
\begin{equation}
c_{3} = \sum^{{\cal{N}}_{SNIa}}_{j = 1}
\frac{1}{\sigma^{2}_{\mathrm{\mu},j}}\,.
\end{equation}
It is worth noting that
\begin{equation}
\chi^{2}_{\mathrm{SN}}(\mu_{0}, \{\theta_{i}\}) = c_{1} - 2 c_{2}
\mu_{0} + c_{3} \mu^{2}_{0} \,,
\end{equation}
which clearly becomes minimum for $\mu_{0} = c_{2}/c_{3}$, so
 that $\tilde{\chi}^{2}_{\mathrm{SN}} \equiv
\chi^{2}_{\mathrm{SN}}(\mu_{0} = c_{2}/c_{3}, \{\theta_{i}\})$.
\subsection{Gamma-ray burst Hubble diagram}
Gamma-ray bursts are visible up to high redshifts thanks to the enormous energy that they release, and thus may be good
candidates  for our high-redshift cosmological investigation.  We performed our analysis using a new updated GRB Hubble diagram data set obtained by  calibrating a $3$-d  $E_{\rm p,i}$-- $E_{\rm iso}$--z relation. Actually, even if recent studies concerning the reliability of the  $E_{\rm p,i}$ --
$E_{\rm iso}$ relation confirmed the  lack, up to now, of  any statistically meaningful evidence for a $z$ dependence
of the correlation coefficients \cite{MGRB1}, we include in the calibration terms representing the $z$-evolution, which
are  assumed to be  power-law functions: $g_{iso}(z)=\left(1+z\right)^{k_{iso}}$ and
$g_{p}(z)=\left(1+z\right)^{k_{p}}$ (see for instance\citep[][]{MGRB1}), so that $E_{\rm iso}^{'}
=\displaystyle\frac{E_{\rm iso}}{g_{iso}(z)}$ and $E_{\rm p,i}^{'} =\displaystyle\frac{E_{\rm p,i}}{g_{p}(z)}$ are the
de-evolved  quantities. Therefore we consider  a 3D correlation:
\begin{eqnarray}
 \label{eqamatievol}
&&\log \left[\frac{E_{\rm iso}}{1\;\mathrm{erg}}\right] = b+a \log  \left[
    \frac{E_{\mathrm{p,i}} }{300\;\mathrm{keV}} \right]+\nonumber \\&& + \left(k_{iso} - a \,k_{p}\right)\log\left(1+z\right)\,.
\end{eqnarray}
 In order to calibrate our de-evolved  relation we  apply  the same  local regression
technique  previously adopted \citep{MGRB1,MGRB2}, but we consider a 3D Reichart
likelihood:
\begin{eqnarray}
 \label{reich3dl}
&&L^{3D}_{Reichart}(a, k_{iso}, k_{p}, b,  \sigma_{int})=\nonumber\\&& \frac{1}{2} \frac{\sum{\log{(\sigma_{int}^2 + \sigma_{y_i}^2 + a^2
\sigma_{x_i}^2)}}}{\log{(1+a^2)}}\nonumber \\ &+& \frac{1}{2} \sum{\frac{(y_i - a x_i -(k_{iso}-\alpha) z_i-b)^2}{\sigma_{int}^2 + \sigma_{x_i}^2 + a^2
\sigma_{x_i}^2}}\,,
\end{eqnarray}
where $\alpha= a\, k_{p}$. We also used the MCMC method to maximize the likelihood and ran
five parallel chains and the Gelman-Rubin convergence test. We
found that $a=1.87^{+0.08}_{-0.09}$, $k_{iso}=-0.04\pm 0.1$;
$\alpha=0.02\pm 0.2$\,; $\sigma_{int}=0.35_{-0.03}^{+0.02}$, so
that $b= 52.8_{-0.06}^{+0.03}$. After fitting the  correlation
and estimating its parameters, we used them to construct the GRB
Hubble diagram. 
\subsection{Baryon acoustic oscillations  data}
Baryon acoustic oscillations data are promising standard rulers to investigate
different cosmological scenarios and models. They  are related to density
fluctuations induced by acoustic waves that are created by primordial
perturbations. To use
BAOs as a cosmological tool, we define:\,

\begin{equation}
d_z = \frac{r_s(z_d)}{d_V(z)}\,,
\label{eq: defdz}
\end{equation}
where $z_d$ is the drag redshift, $r_s(z)$ is the
comoving sound horizon\,,
\begin{equation}
r_s(z) = \frac{c}{\sqrt{3}} \int_{0}^{(1 + z)^{-1}}{\frac{da}{a^2 H(a) \sqrt{1 + (3/4) \Omega_b/\Omega_{\gamma}}}}\ , \nonumber
\label{defsoundhor}
\end{equation}
and $d_V(z)$ the volume distance. Moreover, BAO
measurements in spectroscopic surveys allow to directly estimate
the expansion rate H(z), converted into the quantity
$D_{H}(z)=\displaystyle\frac{c}{H(z)}$, and put constraints  on the comoving angular diameter
distance $D_{M}(z)$. The BAO data used in our analysis are summarized in
Table \ref{tab:data} and are taken from  \cite{Aubourg14}.
\begin{table*}
  \centering
  \setlength{\tabcolsep}{0.4em}
  \begin{tabular}{|cccc|}
  \hline\hline
    Redshift &  $D_V/r_d$         &   $D_M/r_d$ &   $D_H/r_d$  \\
    \hline

                        0.106   &  $3.047 \pm 0.137$ &       --    &  --                       \\

                         0.15     &  $4.480 \pm 0.168$ &        --         & --     \\

    0.32    &  $8.467 \pm 0.167$  &       --     & --   \\
 0.57    &    --                    &  $14.945 \pm 0.210$ & $20.75 \pm 0.73 $  \\

2.34 &    --               &  $37.675 \pm 2.171$ & $9.18 \pm 0.28 $ \\
 2.36 & --             &  $36.288 \pm 1.344$ & $9.00 \pm 0.30 $  \\
    2.34    &    --               &  $36.489 \pm 1.152$ & $9.145 \pm 0.204 $ \\
\hline
  \end{tabular}

  \caption{
    BAO data used in our analysis.
  }
  \label{tab:data}
\end{table*}
Here, the BAO scale $r_d$  is the radius of the sound horizon at the
decoupling era.
\subsection{H(z) measurements}
 The measurements of Hubble parameters are a complementary probe to constrain
the cosmological parameters and investigate the dark energy \cite{farooqb}. The Hubble parameter depends on the differential age of the Universe as a
function of redshift and can be measured  using the so-called cosmic chronometers. $dz$ is obtained from spectroscopic
surveys with high accuracy, and  the differential evolution of the age of the Universe $dt$ in the  redshift interval
$dz$ can be measured provided that  optimal probes of the aging of the Universe, that is, the cosmic chronometers, are
identified. The most reliable cosmic chronometers at present are old early-type galaxies that evolve passively on a
timescale much longer than their age difference, which formed the vast majority of their stars rapidly and  early  and
have not experienced  subsequent major episodes of star formation or merging. Moreover, the Hubble parameter can also
be obtained from the BAO measurements. We used a list of $28$ $H(z)$
measurements, compiled in \cite{farooqb}.
\section{Statistical analysis}
To test the cosmological parameters described above, we use a Bayesian approach based on MCMC method. In order to set
the starting points for our chains, we first performed a preliminary and standard fitting procedure to maximize the
likelihood function ${\cal{L}}({\bf p})$:
\begin{eqnarray}
\footnotesize
{\mathcal{L}}({\bf p}) & \propto & \frac {\exp{(-\chi^2_{SNIa/GRB}/2)}}{(2 \pi)^{\frac{{\cal{N}}_{SNIa/GRB}}{2}} |{\bf C}_{SNIa/GRB}|^{1/2}} \times\nonumber\\&& \frac{\exp{(-\chi^2_{BAO}/2})}{(2 \pi)^{{\cal{N}}_{BAO}/2} |{\bf C}_{BAO}|^{1/2}}\times \nonumber\\ ~ & \times & \frac{1}{\sqrt{2 \pi \sigma_{\omega_m}^2}}
 \exp{\left [ - \frac{1}{2} \left ( \frac{\omega_m - \omega^{obs}_{m}}{\sigma_{\omega_m}} \right )^2\right ]} \\ &&\times\frac{1}{\sqrt{2 \pi \sigma_h^2}} \exp{\left[ - \frac{1}{2} \left ( \frac{h - h_{obs}}{\sigma_h} \right )^2\right]} \nonumber\\ && \frac{\exp{(-\chi^2_{H}/2})}{(2 \pi)^{{\cal{N}}_{H}/2} |{\bf C}_{H}|^{1/2}}\nonumber
 \\ & \times & \frac{1}{\sqrt{2 \pi \sigma_{{\cal{R}}}^2}} \exp{\left [ - \frac{1}{2} \left ( \frac{{\cal{R}} - {\cal{R}}_{obs}}{\sigma_{{\cal{R}}}} \right )^2 \right ]}\,. \nonumber \,
\label{defchiall}
\end{eqnarray}

Here

\begin{equation}
\chi^2(\mathrm {\bf p}) = \sum_{i,j=1}^{N} \left( x_i -
x^{th}_i(\bf p)\right)C^{-1}_{ij}  \left( x_j - x^{th}_j(\bf
p)\right) \,, \label{eq:chisq}
\end{equation}
 $\bf p$ is the set of parameters, $N$ is the number of data points,  $\mathrm x_i$ is the $i-th$ measurement;
$ x^{th}_i(\bf p)$ indicate the theoretical predictions for
these measurements and depend on the parameters $\bf p$.
$C_{ij}$ is the covariance matrix (specifically, ${\bf
C}_{SNIa/GRB/H}$ indicates the SNIa/GRBs/H  covariance matrix);
$(h^{obs}, \sigma_h) = (0.742, 0.036)$\, \citep{shoes}, and
$(\omega_{m}^{obs}, \sigma_{\omega_m}) = (0.1356, 0.0034)$\,
\citep{PlanckXIII}. It is worth noting that  the effect of our
prior on h is not critical at all so that we are certain that
our results are not biased by this choice. The term $\displaystyle  \frac{1}{\sqrt{2 \pi
\sigma_{{\cal{R}}}^2}} \exp{\left [ - \frac{1}{2} \left (
\frac{{\cal{R}} - {\cal{R}}_{obs}}{\sigma_{{\cal{R}}}} \right
)^2 \right ]} $ in the likelihood (\ref{defchiall}) considers
the shift parameter ${\cal{ R}}$:

\begin{equation}
{\cal{R}} = H_{0} \sqrt{\Omega_M} \int_{0}^{z_{\star}}{\frac{dz'}{H(z')}}\,,
\label{eq: defshiftpar}
\end{equation}
where $z_\star = 1090.10$ is the redshift of the surface of last scattering  \citep{B97,EB99}. According to
the Planck data  $({\cal{R}}_{obs}, \sigma_{{\cal{R}}}) = ( 1.7407, 0.0094)$.

Finally, the term  $\displaystyle
\frac{\exp{(-\chi^2_{H}/2})}{(2 \pi)^{{\cal{N}}_{H}/2} |{\bf
C}_{H}|^{1/2}}  $ in Eq. (\ref{defchiall}) is the likelihood
relative to the measurements of $H(z)$ . For each cosmological
model we sample its space of parameters, by  running five
parallel chains and use the Gelman - Rubin diagnostic approach
to test the convergence. In Tables \ref{tabcpl},  \ref{tabscalar}, and \ref{tabede}
we present the results of our analysis.
\begin{table*}
\begin{center}
%\scriptsize
\resizebox{8cm}{!}{
\begin{tabular}{cccccccccc}
\hline
~ & \multicolumn{7}{c}{\bf CPL Parametrization}   \\
~ & ~ & ~ & ~ & ~ & ~ & ~ & ~ & ~   \\
\hline
~ & ~ & ~ & ~ & ~ & ~ & ~ & ~ & ~   \\
$Id$ & $\langle x \rangle$ & $\tilde{x}$ & $68\% \ {\rm CL}$  & $95\% \ {\rm CL}$ &  $\langle x \rangle$ & $\tilde{x}$ & $68\% \ {\rm CL}$  & $95\% \ {\rm CL}$ \\
\hline \hline
~ & ~ & ~ & ~ & ~ & ~ & ~ & ~ & ~   \\
\hline ~ & \multicolumn{4}{c}{Full dataset}  ~ & \multicolumn{4}{c}{No SNIa}
 \\
\hline
~ & ~ & ~ & ~ & ~ & ~ & ~ & ~ & ~   \\
$\Omega_m$ &0.23 &0.24& (0.19, 0.27) & (0.14, 0.29) &0.19 &0.2& (0.16, 0.22) & (0.10, 0.27)  \\
~ & ~ & ~ & ~ & ~ & ~ & ~ & ~ & ~  \\
$\Omega_b$ & 0.046& 0.046 & (0.04, 0.047) & (0.043, 0.049)  & 0.055& 0.054 & (0.045, 0.068) & (0.037, 0.06) \\
~ & ~ & ~ & ~ & ~ & ~ & ~ & ~ & ~ \\
$w_0$ &-0.88& -0.87& (-1.0, -0.74) & (-1.18,  -0.67)  &-0.7& -0.7& (-0.8, -0.62) & (-1.05, -0.6) \\
~ & ~ & ~ & ~ & ~ & ~ & ~ & ~ & ~ &  \\
$w_a$ &0.16& 0.17 & (-0.15, 0.43) & (-0.3, 0.49) &0.39& 0.42& (0.3,0.48) & (0.1, 0.52)  \\
~ & ~ & ~ & ~ & ~ & ~ & ~ & ~ & ~ &  \\
$h$ &0.69& 0.69 & (0.68, 0.71) & (0.67, 0.72)  &0.67& 0.67 & (0.64, 0.69) & (0.64, 0.72)  \\
~ & ~ & ~ & ~ & ~ & ~ & ~ & ~ & ~ &  \\
\hline

\end{tabular}}
\end{center}
\caption{Constraints on the EOS parameters for the CPL model. } \label{tabcpl}
\end{table*}

\begin{table*}
\begin{center}
%\scriptsize
\resizebox{8cm}{!}{
\begin{tabular}{cccccccccc}
\hline
~ & \multicolumn{7}{c}{\bf Scalar field Quintessence}   \\
~ & ~ & ~ & ~ & ~ & ~ & ~ & ~ & ~   \\
\hline
~ & ~ & ~ & ~ & ~ & ~ & ~ & ~ & ~   \\
$Id$ & $\langle x \rangle$ & $\tilde{x}$ & $68\% \ {\rm CL}$  & $95\% \ {\rm CL}$ &  $\langle x \rangle$ & $\tilde{x}$ & $68\% \ {\rm CL}$  & $95\% \ {\rm CL}$ \\
\hline \hline
~ & ~ & ~ & ~ & ~ & ~ & ~ & ~ & ~   \\
\hline ~ & \multicolumn{4}{c}{Full dataset}  ~ & \multicolumn{4}{c}{No SNIa}
 \\
\hline
~ & ~ & ~ & ~ & ~ & ~ & ~ & ~ & ~   \\
$\Omega_b$ & 0.051& 0.051 & (0.050, 0.051) & (0.049, 0.052) &0.051&0.051&(0.050, 0.0514)&(0.049,0.052)\\
~ & ~ & ~ & ~ & ~ & ~ & ~ & ~ & ~ \\
$H_0$ &0.98&0.98&(0.95,0.99)&(0.94, 1.01) &0.96& 0.96& (0.94, 0.98) & (0.92, 1.05)\\
~ & ~ & ~ & ~ & ~ & ~ & ~ & ~ & ~ &  \\
$h$ &0.69&0.68&(0.67,0.695)&(0.67, 0.7) &0.67& 0.67 & (0.65, 0.68) & (0.64, 0.70) \\
~ & ~ & ~ & ~ & ~ & ~ & ~ & ~ & ~ &  \\
\hline
\end{tabular}}
\end{center}
\caption{Constraints on the parameters for the scalar field quintessence model. } \label{tabscalar}
\end{table*}

\begin{table*}
\begin{center}
%\scriptsize
\resizebox{8cm}{!}{
\begin{tabular}{cccccccccc}
\hline
~ & \multicolumn{7}{c}{\bf Early Dark Energy}  \\
~ & ~ & ~ & ~ & ~ & ~ & ~ & ~ & ~   \\
\hline
~ & ~ & ~ & ~ & ~ & ~ & ~ & ~ & ~   \\
$Id$ & $\langle x \rangle$ & $\tilde{x}$ & $68\% \ {\rm CL}$  & $95\% \ {\rm CL}$ &  $\langle x \rangle$ & $\tilde{x}$ & $68\% \ {\rm CL}$  & $95\% \ {\rm CL}$ \\
\hline \hline
~ & ~ & ~ & ~ & ~ & ~ & ~ & ~ & ~   \\
\hline ~ & \multicolumn{4}{c}{Full dataset}  ~ &
\multicolumn{4}{c}{No SNIa}
 \\
\hline
~ & ~ & ~ & ~ & ~ & ~ & ~ & ~ & ~   \\
$\Omega_m$ &0.29 &0.29& (0.27, 0.31) & (0.25, 0.33) &0.285&0.285&(0.271, 0.298 )&(0.258, 0.312) \\
~ & ~ & ~ & ~ & ~ & ~ & ~ & ~ & ~  \\
$\Omega_b$ & 0.047& 0.048 & (0.040 0.05) & (0.037, 0.052)& 0.045& 0.048 & (0.035, 0.047) & (0.032, 0.054) \\
~ & ~ & ~ & ~ & ~ & ~ & ~ & ~ & ~ \\
$w_0$ &-0.66& -0.67& (-0.85, -0.56) & (-1.33, -0.5) &-0.65&-0.63&(-0.75, -0.53)&(-0.85,-0.50)\\
~ & ~ & ~ & ~ & ~ & ~ & ~ & ~ & ~ &  \\
$\Omega_e$ &0.04& 0.035 & (0.032, 0.043) & (0.026, 0.05) &0.025& 0.023&(0.009, 0.039)&(0.023, 0.03)\\
~ & ~ & ~ & ~ & ~ & ~ & ~ & ~ & ~ &  \\
$h$ &0.71& 0.71 & (0.69, 0.71) & (0.69, 0.72)&0.71&0.71&(0.67,0.73)&(0.67, 0.75)  \\
~ & ~ & ~ & ~ & ~ & ~ & ~ & ~ & ~ &  \\
\hline\hline

\end{tabular}}
\end{center}
\caption{Constraints on the  parameters for the EDE model.
} \label{tabede}
\end{table*}
\section{Prospectives with THESEUS}
So far we showed that the  $E_{\rm p,i}$ -- $E_{\rm iso}$ correlation has significant implications  for the use of GRBs in cosmology and therefore GRBs are powerful cosmological probe, complementary to other probes. Future GRB missions, like, e.g., the proposed THESEUS observatory \citep{Amati_theseus}, will increase substantially the number of GRB usable to construct  the $E_{\rm p,i}$ -- $E_{\rm iso}$  correlation  up to redshift $ z \simeq 10$ and will allow a better calibration of the correlation. 
Here, we compare  the confidence intervals on the cosmological parameters for a FLRW flat model, for the CPL parametrization of the dark energy EOS,  obtained with the real sample of our 162 GRBs and a simulated sample of 772 objects. The simulated data set was obtained by taking into account the distribution of the observed $E_{\rm p,i}$ -- $E_{\rm iso}$ correlation, the distribution of the uncertainties in the measured values of$E_{\rm p,i}$ and $E_{\rm iso}$, and the observed redshift distribution of GRBs.  In order to build up our simulated GRB Hubble diagram data set, we  first calibrate our $3$-d  $E_{\rm p,i}$-- $E_{\rm iso}$--z relation. Therefore we start the cosmological investigations, considering only the CPL model. In Table \ref{tabcplsimul} are summarized  the results of our analysis:  with our mock sample of  GRBs the accuracy in measuring $\Omega_m$ and the dark energy EOS will be comparable to that currently provided by SNe data.
\begin{table*}
\begin{center}
%\scriptsize
\resizebox{8cm}{!}{
\begin{tabular}{cccccccccc}
\hline
~ & \multicolumn{7}{c}{\bf CPL Parametrization}   \\
~ & ~ & ~ & ~ & ~ & ~ & ~ & ~ & ~   \\
\hline
~ & ~ & ~ & ~ & ~ & ~ & ~ & ~ & ~   \\
$Id$ & $\langle x \rangle$ & $\tilde{x}$ & $68\% \ {\rm CL}$  & $95\% \ {\rm CL}$ &  $\langle x \rangle$ & $\tilde{x}$ & $68\% \ {\rm CL}$  & $95\% \ {\rm CL}$ \\
\hline \hline
~ & ~ & ~ & ~ & ~ & ~ & ~ & ~ & ~   \\
\hline ~ & \multicolumn{4}{c}{Full dataset}  ~ & \multicolumn{4}{c}{No SNIa}
 \\
\hline
~ & ~ & ~ & ~ & ~ & ~ & ~ & ~ & ~   \\
$\Omega_m$ &0.19 &0.19& (0.16, 0.23) & (0.14, 0.26) &0.17 &0.17& (0.16, 0.2) & (0.11, 0.23)  \\
~ & ~ & ~ & ~ & ~ & ~ & ~ & ~ & ~  \\
$\Omega_b$ & 0.046& 0.046 & (0.04, 0.047) & (0.043, 0.049)  & 0.055& 0.054 & (0.045, 0.068) & (0.037, 0.06) \\
~ & ~ & ~ & ~ & ~ & ~ & ~ & ~ & ~ \\
$w_0$ &-0.8& -0.78& (-0.93, -0.69) & (-1.04,  -0.63)  &-0.7& -0.7& (-0.75, -0.63) & (-0.81, -0.6) \\
~ & ~ & ~ & ~ & ~ & ~ & ~ & ~ & ~ &  \\
$w_a$ &0.32& 0.34& (0.19, 0.44) & (0.05, 0.5) &0.38& 0.39& (0.3,0.47) & (0.12, 0.49)  \\
~ & ~ & ~ & ~ & ~ & ~ & ~ & ~ & ~ &  \\
$h$ &0.66& 0.69 & (0.68, 0.70) & (0.67, 0.71)  &0.67& 0.67 & (0.64, 0.68) & (0.62, 0.72)  \\
~ & ~ & ~ & ~ & ~ & ~ & ~ & ~ & ~ &  \\
\hline

\end{tabular}}
\end{center}
\caption{Constraints on the EOS parameters for the CPL model from the simulated dataset. } \label{tabcplsimul}
\end{table*}

%\begin{figure*}
%\centerline {\includegraphics[width=4 cm,height=4cm]{GRBHDsimul.eps}}
%\caption{\footnotesize Simulated Hubble diagram}
%\label{GRBHDsimul}
%\end{figure*}

\section{Discussion and conclusions}
The  $E_{\rm p,i}$ -- $E_{\rm iso}$ correlation has significant implications  for the use of GRBs in cosmology. Here we
explored a 3D Amati relation in a way independent of the cosmological model, and taking into account a possible
redshift evolution effects of its correlation coefficients \citep{MGRB1} parametrized as power low terms:
$g_{iso}(z)=\left(1+z\right)^{k_{iso}}$ and $g_{p}(z)=\left(1+z\right)^{k_{p}}$. Low values of $k_{iso}$ and $k_{p}$
would indicate negligible evolutionary effects.
 Using the recently updated data set of 162 high-redshift GRBs, we  applied a local regression technique to estimate the distance modulus
using the recent Union SNIa sample (Union2.1). The derived calibration parameters are statistically fully consistent
with the  results of our  previous work  \citep[][]{MEC11,MGRB1}, and confirm that the correlation shows, at this
stage, only weak indication of evolution. The fitted calibration parameters have been used to construct a calibrated
GRB Hubble diagram, which we adopted as a tool  to constrain different cosmological models: we considered the CPL
parameterization of the EOS, an exponential dark energy scalar field, and,  finally a model with dark energy at early
times. To compare these models we  assumed that the CPL is true  and checked the occurrence of
$\chi^2_{EDE/Quintessence} < \chi^2_{CPL}$, varying the parameters specific of the EDE and scalar field model
respectively. It turns out that the EDE and the scalar field quintessence are slightly favored by the present data. Moreover, it is worth noting  that, also without the
SNIa, the GRBs Hubble diagram is able to set the transition region from the decelerated to the accelerated expansion in
all the tested cosmological models. This definitively proves that GRBs are  powerful cosmological probe, complementary to other probes.  
Future GRBs missions (THESEUS) will increase  the number of GRB usable to construct  the 
$E_{\rm p,i}$ -- $E_{\rm iso}$  correlation  up to redshift $ z \simeq 10$ and will allow a better calibration of the correlation. 
Probably also a self-calibration will be available. Therefore,  the effective role of z evolution will be clarified, 
and  the GRBs Hubble diagram will be able to measure the cosmological parameters and to test the evolution 
of dark energy, in a complementary way to type Ia SNe. Indeed, we used a simulated dataset of 772 GRBs to 
constraint  the cosmological parameters for a FLRW flat model, in the case of the CPL parametrization of 
the dark energy EOS:  it turns out that the accuracy in measuring $\Omega_m$, $h$, and the dark energy EOS, 
will be competitive with respect to that  currently provided by SNe data.

\subsection*{Acknowledgments}
MD is grateful to the INFN for financial support through the Fondi FAI GrIV.
EP acknowledges the support of INFN Sez. di Napoli  (Iniziative Specifica QGSKY ).
LA  acknowledges suppo\ rt by the Italian Ministry for Education,
University and Research through PRIN MIUR 2009 project on "Gamma ray
bursts: from progenitors to the physics of the prompt emission process." (Prot. 2009 ERC3HT).

\bibliographystyle{aa}

\end{document}